\documentclass[%
aps,
superscriptaddress,
showpacs,
prc,
twocolumn,
floatfix,
]{revtex4-1}

\usepackage{graphicx}% Include figure files
\usepackage{dcolumn}% Align table columns on decimal point
\usepackage{bm}% bold math
\usepackage[displaymath, mathlines]{lineno}
\usepackage{color}
\usepackage{times}
\usepackage{hyperref}
\usepackage{breakurl}
\newcommand{\SINICAADDRESS}{Institute of Physics, Academia Sinica, Taipei 11529, Taiwan }
\newcommand{\KUADDRESS}{Department of Physics, Korea University, Seoul 02841, Republic of Korea}

\newcommand{\SEOULADDRESS}{Rare Isotope Science Project, Insitute for Basic Science, Daejeon 34047, Korea}
\newcommand{\OHIOADDRESS}{Department of Physics and Astronomy, Ohio University, Athens, Ohio 45701, USA}
\newcommand{\KYOTOADDRESS}{Department of Physics, Kyoto University, Kyoto 606-8502, Japan}

\newcommand{\NCUADDRESS}{Department of Physics, National Central University, Taoyuan City 32001, Taiwan}

\newcommand{\NSRRCADDRESS}{Light Source Division, National Synchrotron Radiation Research Center, Hsinchu, 30076, Taiwan }
\newcommand{\RCNPADDRESS}{Research Center for Nuclear Physics, Osaka University, Ibaraki, Osaka 567-0047, Japan}
\newcommand{\JASRIADDRESS}{Japan Synchrotron Radiation Research Institute, Sayo, Hyogo 679-5143, Japan}

\newcommand{\NAGOYAADDRESS}{Kobayashi-Maskawa Institute, Nagoya University, Nagoya, Aichi 464-8602, Japan}
\newcommand{\TOHOKUADDRESS}{Research Center for Electron Photon Science, Tohoku University, Sendai, Miyagi 982-0826, Japan}

\newcommand{\TOKYOIADDRESS}{Department of Physics, Tokyo Institute of Technology, Tokyo 152-8551, Japan}
\newcommand{\GIFUADDRESS}{Department of Education, Gifu University, Gifu 501-1193, Japan}
\newcommand{\RIKENADDRESS}{RIKEN Nishina Center, 2-1 Hirosawa, Wako, Saitama 351-0198, Japan}

\newcommand{\WAKAYAMAADDRESS}{Wakayama Medical College, Wakayama, 641-8509, Japan}
\newcommand{\SASKAADDRESS}{Department of Physics and Engineering Physics, University of Saskatchewan, Saskatoon, SK S7N 5E2, Canada}
\newcommand{\KEKADDRESS}{High Energy Accelerator Organization (KEK), Tsukuba, Ibaraki 305-0801, Japan}

\newcommand{\CONNEADDRESS}{Department of Physics, University of Connecticut, Storrs, Connecticut 06269-3046, USA}
\newcommand{\FUKUIADDRESS}{Proton Therapy Center, Fukui Prefectural Hospital, Fukui 910-8526, Japan}
\newcommand{\JAEAADDRESS}{Advanced Science Research Center, 
Japan Atomic Energy Agency, Tokai, Ibaraki 319-1195, Japan}

\newcommand{\KRISSADDRESS}{Korea Research Institute of Standards 
and Science (KRISS), Daejeon 34113, Republic of Korea}
\newcommand{\GENKENADDRESS}{National Institutes for Quantum and Radiological Science and Technology, Tokai, Ibaraki 319-1195, Japan}
\newcommand{\MICHIGANADDRESS}{Physics Department, University of Michigan, Michigan 48109-1040, USA}
\newcommand{\BEIJINGADDRESS}{Institute of High Energy Physics, Chinese Academy of Sciences, Beijing 100049, China}
\newcommand{\TOKYOHADDRESS}{Department of Radiology, The University of Tokyo Hospital, Tokyo 113-8655, Japan}
\newcommand{\CROSSADDRESS}{Neutron Science and Technology Center, Comprehensive Research Organization for Science and Society (CROSS), Tokai, Ibaraki 319-1106, Japan}
\newcommand{\GRADUATEADDRESS}{Graduate School of Science, Kyoto University, Kyoto 606-8502, Japan}

\newcommand{\cmangle} {$\cos \theta_{c.m.}^{K^+}$}
\newcommand{\eg} {$E_{\gamma}$}

\begin{document}

%\linenumbers

%\preprint{APS/123-QED}

\title{Photoproduction of $\Lambda$ and $\Sigma^{0}$ hyperons off
  protons with linearly polarized photons at \\
$E_{\gamma} =$ 1.5--3.0 GeV}

\author{S.~H.~Shiu}\affiliation{\SINICAADDRESS}\affiliation{\NCUADDRESS}
\author{H.~Kohri}\affiliation{\RCNPADDRESS}\affiliation{\SINICAADDRESS}
\author{W.~C.~Chang}\affiliation{\SINICAADDRESS}
\author{D.~S.~Ahn}\affiliation{\RIKENADDRESS}
\author{J.~K.~Ahn}\affiliation{\KUADDRESS}
\author{J.~Y.~Chen}\affiliation{\NSRRCADDRESS}
\author{S.~Dat$\acute{\rm{e}}$}\affiliation{\JASRIADDRESS}
\author{H.~Ejiri}\affiliation{\RCNPADDRESS}
\author{H.~Fujimura}\affiliation{\WAKAYAMAADDRESS}
\author{M.~Fujiwara}\affiliation{\RCNPADDRESS}\affiliation{\GENKENADDRESS}
\author{S.~Fukui}\affiliation{\RCNPADDRESS}
\author{W.~Gohn}\affiliation{\CONNEADDRESS}
\author{K.~Hicks}\affiliation{\OHIOADDRESS}
\author{T.~Hotta}\affiliation{\RCNPADDRESS}
\author{S.~H.~Hwang}\affiliation{\KRISSADDRESS}
\author{K.~Imai}\affiliation{\JAEAADDRESS}
\author{T.~Ishikawa}\affiliation{\TOHOKUADDRESS}
\author{K.~Joo}\affiliation{\CONNEADDRESS}
\author{Y.~Kato}\affiliation{\NAGOYAADDRESS}
\author{Y.~Kon}\affiliation{\RCNPADDRESS}
\author{H.~S.~Lee}\affiliation{\SEOULADDRESS}
\author{Y.~Maeda}\affiliation{\FUKUIADDRESS}
\author{T.~Mibe}\affiliation{\KEKADDRESS}
\author{M.~Miyabe}\affiliation{\TOHOKUADDRESS}
\author{K.~Mizutani}\affiliation{\KYOTOADDRESS}
\author{Y.~Morino}\affiliation{\KEKADDRESS}
\author{N.~Muramatsu}\affiliation{\TOHOKUADDRESS}
\author{T.~Nakano}\affiliation{\RCNPADDRESS}
\author{Y.~Nakatsugawa}\affiliation{\BEIJINGADDRESS}
\author{M.~Niiyama}\affiliation{\KYOTOADDRESS}
\author{H.~Noumi}\affiliation{\RCNPADDRESS}
%\author{Y.~Oh}\affiliation{\KYUNGADDRESS}
\author{Y.~Ohashi}\affiliation{\JASRIADDRESS}
\author{T.~Ohta}\affiliation{\TOKYOHADDRESS}
\author{M.~Oka}\affiliation{\RCNPADDRESS}
\author{J.~D.~Parker}\affiliation{\CROSSADDRESS}
\author{C.~Rangacharyulu}\affiliation{\SASKAADDRESS}
\author{S.~Y.~Ryu}\affiliation{\RCNPADDRESS}
\author{T.~Sawada}\affiliation{\MICHIGANADDRESS}
\author{H.~Shimizu}\affiliation{\TOHOKUADDRESS}
\author{Y.~Sugaya}\affiliation{\RCNPADDRESS}
\author{M.~Sumihama}\affiliation{\GIFUADDRESS}
\author{T.~Tsunemi}\affiliation{\GRADUATEADDRESS}
\author{M.~Uchida}\affiliation{\TOKYOIADDRESS}
\author{M.~Ungaro}\affiliation{\CONNEADDRESS}
\author{M.~Yosoi}\affiliation{\RCNPADDRESS}

\collaboration{LEPS Collaboration}%\noaffiliation

% but any date may be explicitly specified

\begin{abstract}

We report the measurement of the $\gamma p \rightarrow
K^{+}\Lambda$ and $\gamma p \rightarrow K^{+}\Sigma^{0}$ reactions at SPring-8.
The differential cross sections and photon-beam asymmetries 
are measured at forward $K^{+}$ production angles using 
linearly polarized tagged-photon beams in the range of $E_{\gamma}=1.5$--3.0 GeV.
With increasing photon energy, the cross sections for both
$\gamma p \rightarrow K^{+}\Lambda$ and $\gamma p \rightarrow K^{+}\Sigma^{0}$ reactions decrease slowly.
Distinct narrow structures in the production cross section have not been found at $E_{\gamma}=1.5$--3.0 GeV. 
The forward peaking in the angular distributions of cross sections, a characteristic feature of $t$-channel exchange, is observed for the production of $\Lambda$ in the whole observed energy range.
A lack of similar feature for $\Sigma^{0}$ production reflects a less dominant role of $t$-channel contribution in this channel.
The photon-beam asymmetries remain positive for both reactions, suggesting the dominance of $K^{*}$ exchange in the $t$ channel. 
These asymmetries increase gradually with the photon energy, 
and have a maximum value of +0.6 for both reactions.  
Comparison with theoretical predictions 
based on the Regge trajectory in the $t$ channel and the contributions of 
nucleon resonances indicates the major role of $t$-channel 
contributions as well as non-negligible effects of nucleon resonances
in accounting for the reaction mechanism of hyperon photoproduction in this 
photon energy regime.

\end{abstract}

%\pacs{13.60.Le, 13.60.Rj, 13.88.+e, 14.20.Jn, 25.20.Lj}

\maketitle

%\tableofcontents

\section{Introduction}
\label{sec1}

A quantitative understanding of hadronic interactions at low energies 
has been a long-time challenge.
There exist serious difficulties in deriving hadronic interactions from
the first-principle QCD because of its intrinsic nonperturbative property
in the low-energy regime. 
Instead, an alternative approach is the usage of effective theory where the effective 
Lagrangian is constructed as a sum of all tree-level Feynman diagrams 
in the $s$-, $t$- and $u$-channel exchanges of possible hadrons in 
their ground and excited states. 
Through the comparison of the predictions with the experimental results 
of the differential production cross sections and polarization observables
over a wide-range of hadronic reactions using a variety of beams and targets, 
the hadronic degrees of freedom involved are explored. 
Recent progress is summarized in Refs.~\cite{Mart:2009nj,Klempt:2017lwq}. 
Nevertheless, the identified baryon resonances in the experiments 
and lattice QCD calculations~\cite{Edwards:2011jj} are significantly fewer than what have been predicted by the constituent quark model. 
More experimental and theoretical effort is required to shed light 
on this ``missing'' resonance problem~\cite{Crede:2013sze}. 

The process of hyperon ($Y$) photoproduction has been studied for
exploring the nucleon resonances which couple more strongly to $K Y$ than
to $\pi N$. Furthermore, the isospin structure of $KY$ final states
filters out some possible intermediate states of
nucleon resonances. For example, because of the different isospin
($I$) properties of $\Lambda$ ($I=0$) and $\Sigma^{0}$ ($I=1$), only
$I=1/2$ $N^*$ intermediate states could couple to $K^{+}\Lambda$
while both $I=1/2$ $N^*$ and $I=3/2$ $\Delta^*$ intermediate states
are allowed for $K^{+}\Sigma^{0}$ production.

At \eg~$>2$ GeV, above the resonance region, the $t$-channel
exchange of strange mesons like $K$, $K^*$, and $K_1$ is expected to
play an important role at forward angles. The coupling strength of
the exchanged strange mesons with the ground-state nucleon can be
determined from their $t$-channel contributions.
Measurements of the photon-beam asymmetry help to
further define the hadron photoproduction mechanism
because of the extreme sensitivity to the model
parameters and the presence of resonances.
For example, the measurements of the photon-beam asymmetries
provide unique information to constrain the possible $t$-channel
exchanges. A photon-beam asymmetry close to
$-1$ is expected for the case of dominating unnatural parity
exchange of $K$ or $K_1$ whereas the dominance of natural
parity $K^*$ exchange leads to a photon-beam asymmetry of $+1$.

For the photoproduction of ground-state hyperons $\Lambda$ and
$\Sigma^{0}$ off protons, the measurements of differential cross
sections and various polarization observables of $\gamma p \rightarrow
K^{+}\Lambda$ and $K^{+}\Sigma^{0}$ from threshold up to the photon beam
energy $E_{\gamma}=16$ GeV have been done by
SLAC~\cite{Boyarski:1969iy,Quinn:1979zp},
SAPHIR~\cite{SAPHIR_tran,SAPHIR_glander},
LEPS~\cite{LEPS_zegers,LEPS_sumihama,LEPS_hicks,LEPS_kohri},
CLAS~\cite{CLAS_bradford,CLAS_mccracken,CLAS_dey},
GRAAL~\cite{GRAAL_2007,GRAAL_2009}, and Crystal Ball~\cite{CB_Jude}
experiments. A clear resonance structure in the production cross
  sections around a center-of-mass energy $\sqrt{s}=1.9$--1.96 GeV was observed in the
  $K^{+}\Lambda$ channel~\cite{SAPHIR_tran,SAPHIR_glander,LEPS_sumihama,CLAS_bradford,CLAS_mccracken}, and small enhancement was found at
  $\sqrt{s}=2.05$ GeV at forward angles in the $K^{+}\Sigma^{0}$
  channel~\cite{SAPHIR_tran,SAPHIR_glander,LEPS_sumihama,CLAS_bradford,CLAS_dey}. The photon-beam asymmetries for both channels were seen
  to be positive, and that of $K^{+}\Sigma^{0}$ production was in
  general the larger one of the two. In addition, similar reaction channels for
  $\gamma n \rightarrow K^{+}\Sigma^{-}$ and $\gamma p \rightarrow
  K^{0}\Sigma^{+}$ have also been identified experimentally~\cite{LEPS_kohri,
    CLAS_Pereira, SAPHIR_goers,SAPHIR_lawall,TAPS_2008,TAPS_2012,A2}.

Several theoretical analyses have been performed. 
A common approach is the single- or coupled-channel isobar 
models, e.g., Kaon-MAID 
(KMAID)~\cite{KMAID_1999,KMAID_2014,KMAID_2015,KMAID_2017}. 
In general, a great amount of parameters are required 
in modeling of all 
possible $s$-, $t$-, and $u$-channel Feynman diagrams. 
This introduces 
difficulties in determining such parameters reliably from a limited number of data points available. 

On the other hand, Regge theory is well known, with an elegant 
formalism in modeling high-spin and high-mass particle exchange at 
large $s$ and small $|t|$ or $|u|$. It is very successful in 
describing the diffractive production at high energies. 
The applicability of Regge theory in the low-energy regime is, 
however, controversial. 
Recently, people started to adopt Regge theory 
for modeling the $t$-channel contribution in an effective theory by 
replacing the usual pole-like Feynman propagator with a corresponding 
Regge propagator. 
The number of free parameters was 
significantly reduced. In Ref.~\cite{Guidal:1997hy}, it was found that 
the photoproduction of $\Lambda$ and $\Sigma^{0}$ for $E_{\gamma}=5$--16 GeV 
could be well described by a modified $t$-channel exchange of $K$ and $K^{\ast}$ Regge 
trajectories, together with a corresponding modification of 
the $s$-channel nucleon pole contribution. 
Later, the same approach was successfully extended to describe 
the data in the energy region down to $E_{\gamma}=2$ GeV~\cite{Guidal:2003qs}. 
Therefore, it becomes a popular approach to use 
$K$ and $K^{\ast}$ Regge trajectories in modeling 
the $t$-channel contributions in hyperon photoproduction.

In the Bonn-Gatchina (BG) 
model~\cite{BG_2005,BG_2007,BG_2010,BG_2011,BG_2014}, the strength of 
Regge theory $t$-channel terms is determined simultaneously with the resonance 
contributions in $s$ channels through the data fitting. 
In contrast, the 
Regge-plus-resonance (RPR) model~\cite{RPR_2006,RPR_2007,RPR_2012} 
treats the $t$ channel as a background and fixes its contribution in 
advance by high-energy data at forward production 
angles~\cite{RPR_2007,RPR_2012}, where a forward-peaking 
behavior in the angular distributions is clearly observed. 
The resonance contributions in 
$s$ channels are then added for extrapolation to the resonance 
regions. 
Through these studies, the evidence of a number of nucleon resonances contributing to hyperon photoproduction has been reported~\cite{KMAID_2015,BG_2010,RPR_2012,Anisovich:2017pmi},
e.g. $S_{11}(1650)$, $P_{11}(1710)$, $P_{13}(1720)$ and $D_{13}(1900)$
for $N^*$, and $S_{31}(1900)$, $P_{31}(1910)$,
$D_{33}(1700)$ and $P_{33}(1920)$ for $\Delta^*$.

In the present work, we report on the measurement of the differential
cross sections and photon-beam asymmetries in the energy region of $E_{\gamma}=1.5$--3.0 GeV at very
forward angles. The new results fill the gap at
  $E_{\gamma}=2.5$--6.0 GeV in the existing measurements, providing a
  strong constraint in modeling Regge trajectories in the
  $t$ channel at lower energies, and helping to pin down the contributions
  of heavier nucleon resonances in the transition region. The present paper
is outlined as follows. In Sec.~\ref{sec2}, the
experimental setup and the analysis methods are introduced.  
The results and a comparison with the theoretical
predictions are presented in Sec.~\ref{sec3}.  A summary is given in
Sec.~\ref{sec4}.

%%%
\section{Experiment and data analysis}
\label{sec2}

The experiment was carried out in the laser electron photon beamline at SPring-8 (LEPS) facility in Japan. 
The photon beam was 
generated by the laser backscattering technique using a deep-UV laser 
with a wavelength of 257 nm~\cite{muramatsu}. The energy range of 
the tagged photon beams is 1.5--3.0 GeV, corresponding to $\sqrt{s}=1.92$--2.53 GeV. 
The degree of linear polarization 
for the tagged photon beams is 88\% at $E_{\gamma}=3.0$ GeV, and drops down to 
28\% at $E_{\gamma}=1.5$ GeV. 

\begin{figure}[htb]
\includegraphics[width=0.5\textwidth]{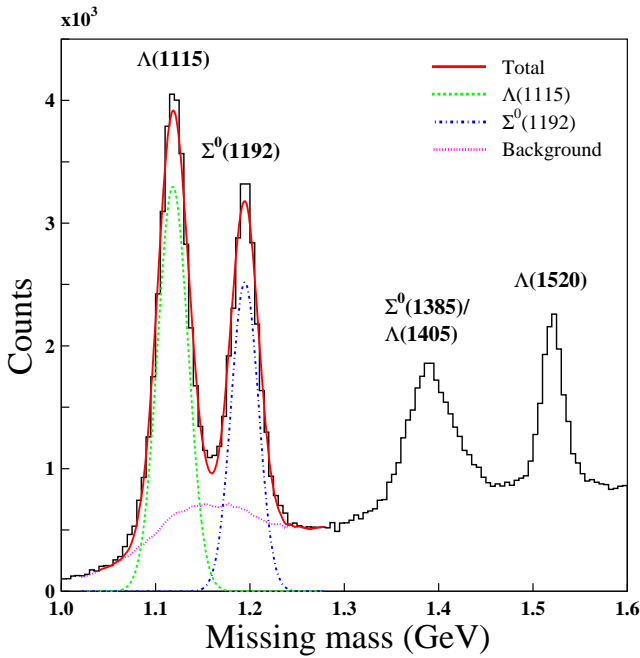}
\caption{\label{fig:miss} Missing mass spectrum of $\gamma p
  \rightarrow K^{+}X$ reaction [$\textrm{MM}_X(\gamma p, K^+)$] at
  $E_{\gamma}=1.5$--3.0 GeV. }
\end{figure}

The photon beam was incident on a liquid hydrogen target with an 
effective length of 16 cm. 
Charged particles produced at the target 
were detected by the LEPS spectrometer in the very forward direction;
the angular coverage is about $\pm 0.4$ and $\pm 0.2$ rad in the 
horizontal and vertical directions, respectively. 
The spectrometer consisted of a dipole magnet, a silicon vertex detector, 
and three drift chambers. 
A time of flight (TOF) measurement for charged particles 
was done using a start counter (SC) scintillator 5 cm downstream of the 
target together with the TOF wall, an array of scintillator bars placed 4 
m downstream of the target. At the end of the beam pipe,
an upstream veto (UPveto) counter made by a plastic scintillator to
eliminate the $e^+e^-$ background events has been installed.
The event trigger was a coincidence of the tagger, UPveto, SC, and TOF wall.
Since the data set was originally 
collected for detecting $K^{\ast 0}$ decaying to high-momentum $K^{+}$ 
and $\pi^{-}$~\cite{hwang}, the regular spectrometer setup was slightly modified. We removed the Aerogel Cherenkov counter, which 
had been placed immediately after the target in earlier experiments to reject high-momentum 
electrons, positrons, and pions at the trigger level. 
The signal from a plastic scintillation counter placed downstream of the drift 
chambers was used to veto $e^+e^-$ pairs 
produced from photon conversion. 
For further details concerning 
the detector configuration and the quality of particle identification, 
refer to Refs.~\cite{nakano,LEPS_sumihama}. 

Particle identification (PID) of the charged particles was done by a 3$\sigma$ 
cut on their reconstructed masses based on the measured TOF, 
momentum and path length, where $\sigma$ is the momentum-dependent mass resolution. 
Pions and kaons were well separated in the 
momenta region lower than 1.0 GeV/$c$.
To ensure good PID by removing $K^{+}$ decay-in-flight,
an extrapolation of the hit position (using the Runge-Kutta
method) from the drift chamber to the TOF wall was applied.
We required the extrapolated vertical hit position to be within 8 cm
of the estimated one based on the time difference of
TOF readout from both ends of the fired scintillator bar.
For the horizontal hit position,
the difference of the channel number of the extrapolated TOF slat and the fired one was required to be less than 2.

\begin{figure}[t]
\includegraphics[width=0.5\textwidth]{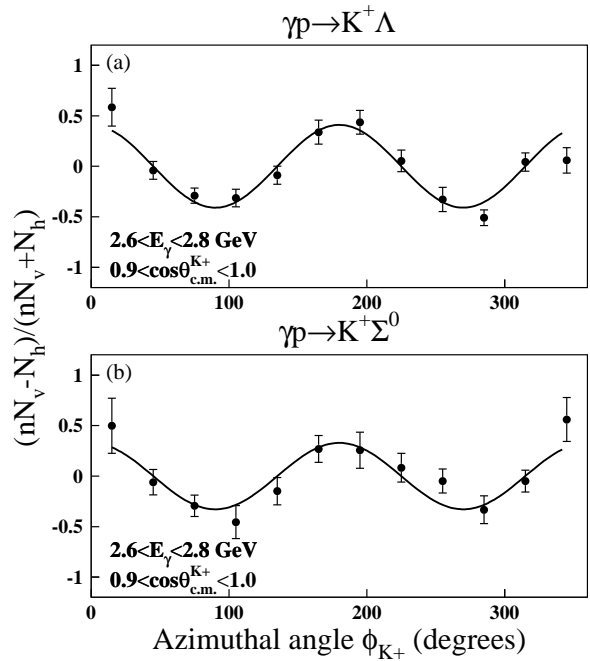}
\caption{\label{fig:bafitex} Azimuthal angle ($\phi_{K^{+}}$)
  dependence of the ratio $(n N_{\rm V}-N_{\rm H})/(n N_{\rm V}+N_{\rm
    H})$ in Eq.~(\ref{eq1}) for the (a) $K^+\Lambda$ and (b) 
  $K^+\Sigma^{0}$ channels at $E_{\gamma} = 2.7$ GeV and 
  $\cos \theta_{c.m.}^{K^+}= 0.9$. The solid lines are the fit results using a function 
  of $\cos 2\phi_{K^{+}}$.}
\end{figure}

With the cuts of an identified $K^{+}$ within 3${\sigma}$ mass 
resolution, a minimum mass of 0.4 GeV/$c^2$, and a vertex position on the 
liquid hydrogen target, the missing mass spectrum for the reaction 
$\gamma p \rightarrow K^{+}X$ [$\textrm{MM}_X(\gamma p, K^+)$] is shown in 
Fig.~\ref{fig:miss} where the $K^{+}\Lambda$ and $K^{+}\Sigma^{0}$ 
events are clearly observed as well as the production of higher 
hyperon resonances such as $\Sigma^0(1385)/\Lambda(1405)$ and 
$\Lambda(1520)$. The numbers of  $K^{+}\Lambda$ and $K^{+}\Sigma^{0}$ events are about $2.6 \times 10^{4}$ and $1.8 \times 10^{4}$, respectively.

Compared to the previous LEPS 
experiment~\cite{LEPS_sumihama}, the background level under the 
 $\Lambda$ and $\Sigma^0$ peaks in the missing mass spectrum 
$\textrm{MM}_X(\gamma p, K^+)$ was enhanced in the current study. 
This is due to the removal 
of the Aerogel Cherenkov counter as mentioned above, and there happened 
to be contamination of pions in the selected $K^{+}$ with momentum
higher than 1.0 GeV/$c$. 
The degree of pion contamination in the selected kaons increased
for particles of larger momenta.
Therefore, the 
fraction of background in the missing mass spectra was enhanced at 
larger \eg~and at smaller kaon production angles. 

\begin{figure*}[t]
\includegraphics[width=0.8\textwidth]{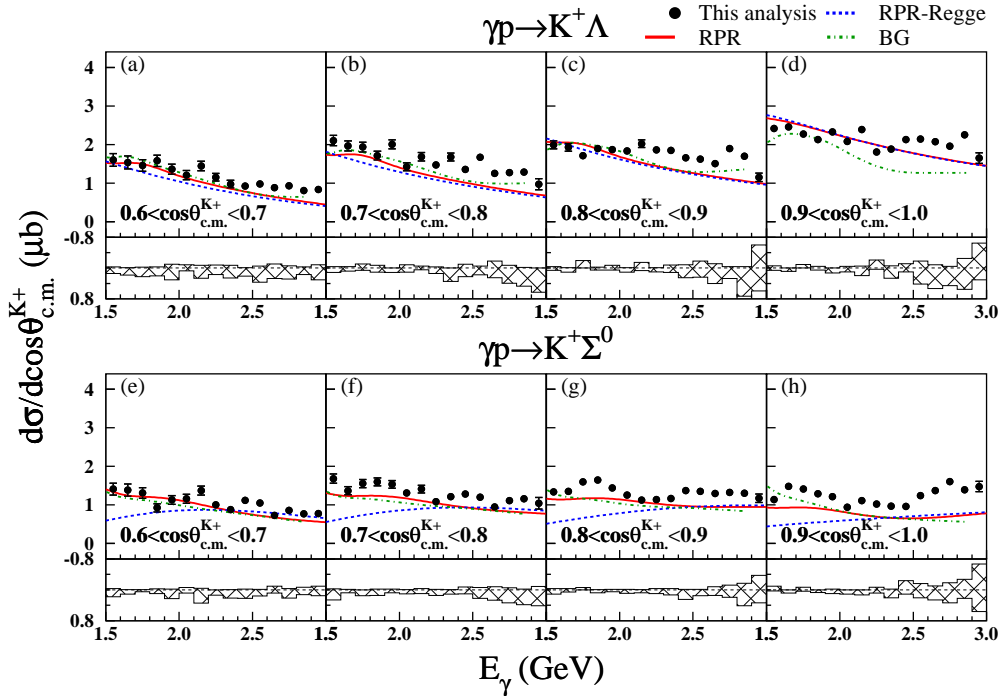}
\caption{\label{fig:creg_merge_4} Differential cross sections for the
  $p$($\gamma$,$K^{+}$)$\Lambda$ reaction (a)--(d) and the $p$($\gamma$,$K^{+}$)$\Sigma^{0}$ 
  reaction (e)--(h) as a function of photon energy $E_{\gamma}$ for the kaon
c.m. production polar angle $0.6<\cos \theta_{c.m.}^{K^+}<1.0$.
The curves denote the predictions of Regge-model~\cite{RPR_url} calculation
with (RPR) and without resonances (RPR-Regge), and Bonn-Gatchina model (BG).
The hatched histograms indicate the systematic uncertainty.}
\end{figure*}

We used a side-band subtraction for eliminating background events caused by the 
misidentified $\pi^{+}$. 
A side-band sample was chosen by using
$\pi^{+}$ events lying outside of the $K^{+}$ mass region in the same bin of track 
momentum ($|\vec{P}|$), photon energy ($E_{\gamma}$), and production 
angle (\cmangle). 
Using the momentum information of 
$\pi^{+}$ particles in the side-band sample, the background template in 
$\textrm{MM}_X(\gamma p, K^+)$ was constructed by 
assuming the kaon mass for the pion tracks. 
With this background template, 
a Monte Carlo model simulated the $\textrm{MM}_X(\gamma p, K^+)$ for $\Lambda$ and 
$\Sigma^{0}$ production in the same \{$|\vec{P}|$, $E_{\gamma}$, 
\cmangle\} bin. A normalization of the background template was obtained 
by fitting the experimental $\textrm{MM}_X(\gamma p, K^+)$ data in the mass range of 
1.0--1.26 GeV/$c^2$. 
The corresponding ranges of $|\vec{P}|$, 
$E_{\gamma}$, and \cmangle\, kinematic variables are 0--3 GeV/$c$, 1.5--3.0 GeV and 0.6--1.0, 
respectively. 
After fixing the normalization in each bin, we summed up the 
background template over all the track momentum ($|\vec{P}|$) bins to 
obtain the combined background template in each \{$E_{\gamma}$, 
\cmangle \} bin. The yields of the $K^{+}\Lambda$ and 
$K^{+}\Sigma^{0}$ production were extracted using another new fit of the 
missing mass spectrum, with two Gaussian distributions 
having constant peaks at standard PDG masses and free widths for the signals, and the combined background template.

Figure~\ref{fig:miss} shows that the missing mass spectrum around 
the signal region of the $K^{+}\Lambda$ and $K^{+}\Sigma^{0}$ events is
well described using two Gaussian distributions of signal (green 
dotted line for $K^{+}\Lambda$ and blue one for $K^{+}\Sigma^{0}$) and 
a background template from the side-band of the kaon mass regions 
(purple dotted line).
The broad bump structure in the background under the $\Lambda$ and $\Sigma^{0}$ peaks is caused by the misidentification of $\pi^{+}$ in the $\gamma p$ $\rightarrow$ $\pi^{+}\Delta^{0}$ reaction.
Another possible background source is the photoproduction of $\phi$ mesons and their charged kaon 
decays. The missing mass distribution of $\phi$ production is expected to appear above 1.5 GeV/$c^2$ in Fig.~\ref{fig:miss}. Therefore it 
does not constitute a background in the signal region of interest for the 
$\Lambda$ and $\Sigma^{0}$ productions. 

In each kinematic bin of $E_{\gamma}$ and \cmangle, the cross sections 
for $K^{+}\Lambda$ and $K^{+}\Sigma^{0}$ photoproduction were evaluated
using the measured yields, the integrated photon flux from the tagger, 
the liquid target density, correction factors for the $K^{+}$ detection, 
and the photon tagging inefficiencies. 
The $K^{+}$ detection efficiency was estimated based 
on Monte Carlo simulations by assuming a uniform production of 
$K^{+}\Lambda$ and $K^{+}\Sigma^{0}$ in $E_{\gamma}$ and \cmangle. 
The acceptance of the LEPS spectrometer was simulated using the \textsc{geant} 
software. 
The simulated peak widths for the missing mass of $K^+$ on 
$\Lambda$ and $\Sigma^{0}$ were in good agreement with those shown in 
Fig.~\ref{fig:miss}.

\begin{figure*}[t]
  \includegraphics[width=0.8\textwidth]{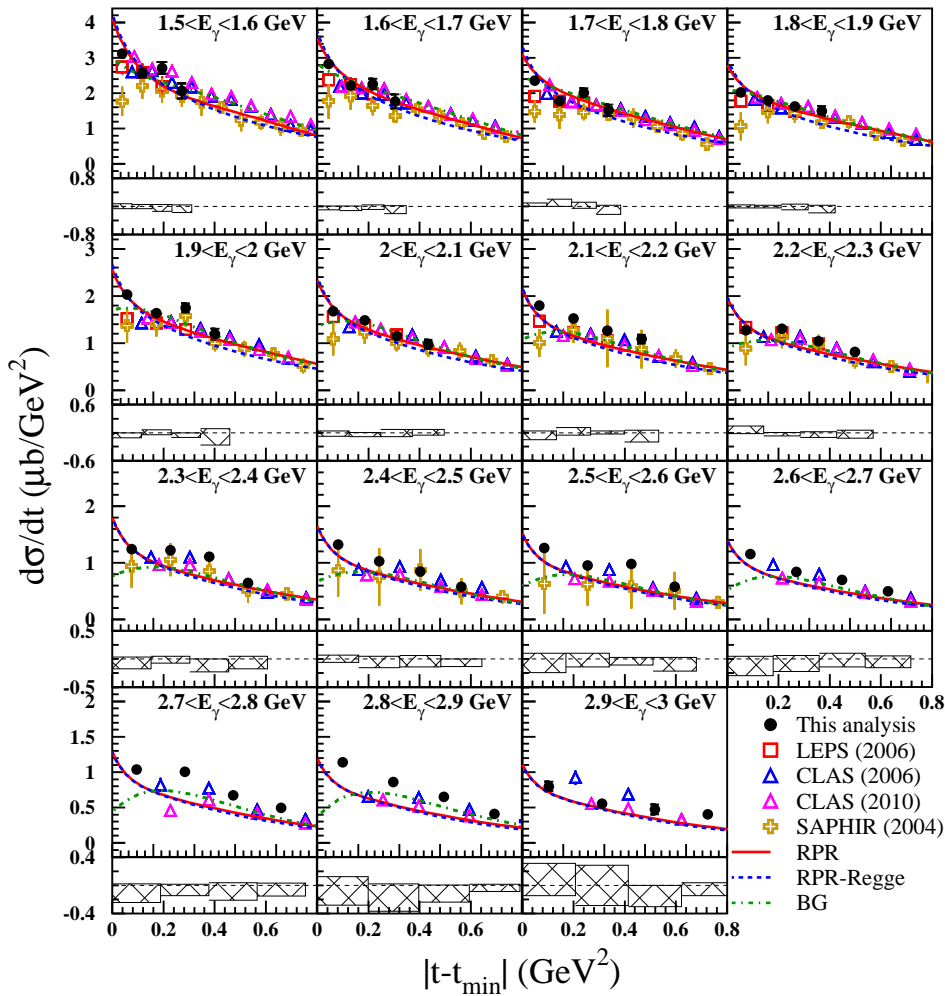}
  \caption{\label{fig:crcos_L_4} Differential cross sections for
    the $p$($\gamma$,$K^{+}$)$\Lambda$ reaction as a function of $|t-t_{min}|$
    for $1.5<E_{\gamma}<3.0$ GeV.
    The results of this measurement are shown by solid black circles.
    The results of LEPS 2006~\cite{LEPS_sumihama},
    CLAS 2006~\cite{CLAS_bradford},
    CLAS 2010~\cite{CLAS_mccracken},
    and SAPHIR 2004~\cite{SAPHIR_glander} are also shown.
    The notations of curves are the same as those in Fig.~\ref{fig:creg_merge_4}.
    The shaded histograms show the systematic uncertainty.
}
\end{figure*}

The systematic uncertainty was estimated by a variation of 
the background template. 
Since there were impurities of $e^{+}$ and 
$K^{+}$ particles in the selected $\pi^{+}$ region used for the 
background template, we observed a dependence of the background template 
on different choices of the selected $\pi^{+}$ region as well as the momentum 
binning. 
The variation range of results, corresponding to the changes 
on the above two factors, was assigned as the major systematic 
uncertainty. 
There were additional systematic uncertainties due to the 
photon-beam flux and the target length which were estimated to be 3\% and 
1\%, respectively.

Using both event yields with the vertically ($\rm V$) and horizontally 
($\rm H$) polarized photon beams, the photon-beam asymmetry
($\Sigma_{\gamma}$) of $K^{+}\Lambda$ and $K^{+}\Sigma^{0}$
photoproduction is given as follows:
\begin{equation}
P_{\gamma}\Sigma_{\gamma} \cos(2\phi_{K^{+}})=\frac{n N_{\rm
    V}-N_{\rm H}}{n N_{\rm V}+N_{\rm H}},
\label{eq1}
\end{equation}
where $P_{\gamma}$ stands for the polarization degree of the photon beams 
and $\phi_{K^{+}}$ denotes the azimuthal production angle of detected 
$K^{+}$ with respect to the horizontal plane in the laboratory system. 
The $N_{\rm V}$ and $N_{\rm H}$ are the individual hyperon production yields
from the vertically and horizontally polarized photon beams, and $n$ is the 
normalization factor defined by the photon flux ratio of 
two polarization directions ($n=n^{\gamma}_{\rm H}/n^{\gamma}_{\rm
  V}$), which is 1.014 for this study. 
Figure~\ref{fig:bafitex} shows 
the ratio $(n N_{\rm V}-N_{\rm H})/(n N_{\rm V}+N_{\rm H})$ as a 
function of $\phi_{K^{+}}$ for the $K^+\Lambda$ and $K^+\Sigma^{0}$ 
channels at $E_{\gamma}=2.7$ GeV and $\cos \theta_{c.m.}^{K^+}=0.9$. 
The ratios are largest at $\phi_{K^{+}}$ = 0$^{\circ}$, 180$^{\circ}$, and 360$^{\circ}$, and
smallest at $\phi_{K^{+}}$ = 90$^{\circ}$ and 270$^{\circ}$ for both the channels. This means
that $K^{+}$ mesons prefer to scatter at $\phi_{K^{+}}$ angles perpendicular
to the polarization plane, suggesting positive photon-beam asymmetries.
The lines overlaid are the best fit of a $\cos(2\phi_{K^{+}})$ modulation.

%%%
\section{Results}
\label{sec3}

\subsection{Differential cross sections}
\label{sec3.1}
The differential cross sections for the $K^{+}\Lambda$ and 
$K^{+}\Sigma^{0}$ reactions as a function of photon-beam energy 
$E_{\gamma}$ in the range of $0.6<\cos \theta_{c.m.}^{K^+}<1.0$ are shown in 
Fig.~\ref{fig:creg_merge_4}. 
The error bars represent the statistical 
errors while the hatched area expresses the range of systematic uncertainty.
The theoretical predictions from the RPR model~\cite{RPR_url} 
with (RPR, solid red lines) and without resonances (RPR-Regge, dashed 
blue lines) as well as BG2014-02 solutions~\cite{BG_url} of 
Bonn-Gatchina (BG) models (dot-dashed green lines) are overlaid for 
comparison. 
We use RPR-2011 solutions~\cite{RPR_2012} 
for $K^{+}\Lambda$ and RPR-2007 ones~\cite{RPR_2007} for 
$K^{+}\Sigma^{0}$.

\begin{figure*}[t]
\includegraphics[width=0.8\textwidth]{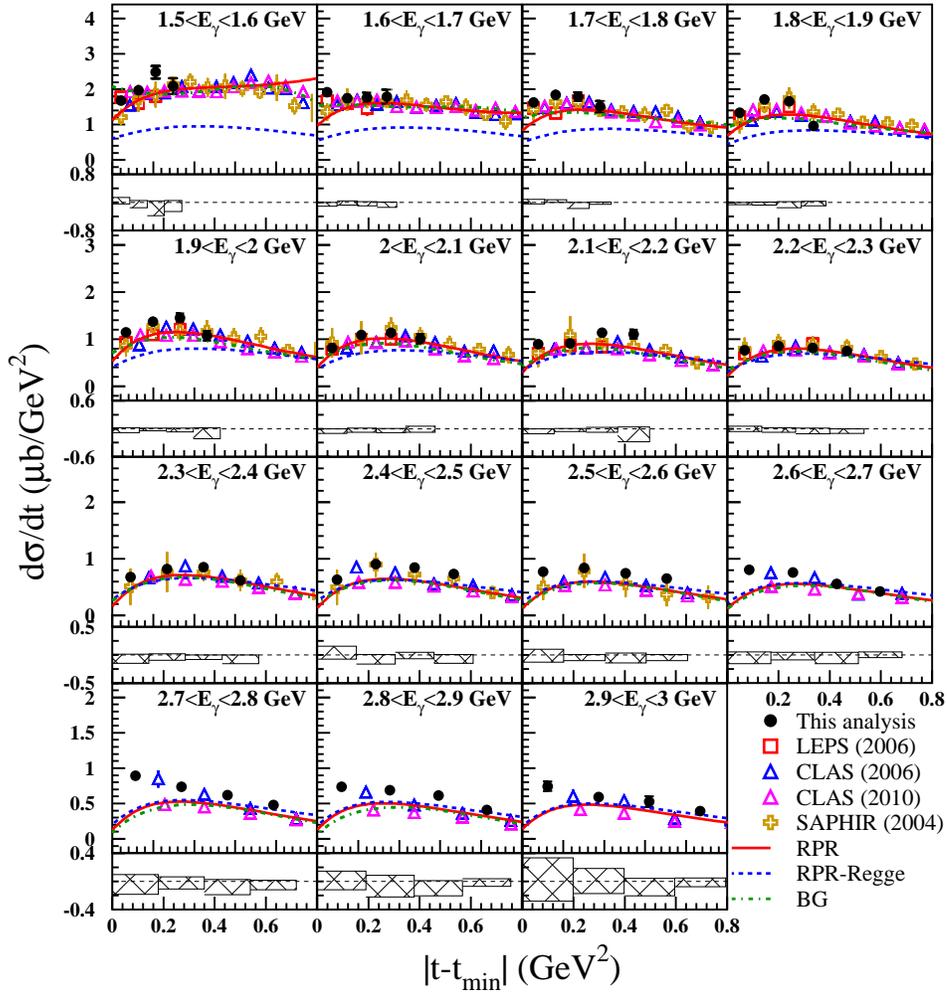}
\caption{\label{fig:crcos_S_4} Same as Fig.~\ref{fig:crcos_L_4}, but for the $p$($\gamma$,$K^{+}$)$\Sigma^{0}$ reaction.
}
\end{figure*}

Within $E_{\gamma} = 1.5$--3.0 GeV, the differential cross sections for 
the $K^{+}{\Lambda}$ channel decrease monotonically with
the beam energy in all four bins of $0.6<\cos \theta_{c.m.}^{K^+}<1.0$. 
No distinct narrow resonance structure is observed.
In contrast, the previous LEPS analysis~\cite{LEPS_sumihama} suggests the observation of a small bump structure around
$E_{\gamma} = 1.5$--1.6 GeV ($\sqrt{s} = 1.92$--1.97 GeV). In this analysis, the larger photon energy bin size prevents
us from identifying the structure.
The cross sections for the $K^{+}{\Lambda}$ channel 
are larger than those for the $K^{+}\Sigma^{0}$ channel. 
The cross sections for the $K^{+}{\Sigma^{0}}$ channel also 
show a decreasing trend with the beam energy whereas its
  energy dependence is relatively mild compared to 
the $K^{+}{\Lambda}$ channel. The mild energy dependence for the $K^{+}\Sigma^{0}$ channel 
is consistent with non-negligible $s$-channel contributions.
Although there is no distinct narrow resonance structure in the 
$K^{+}{\Sigma^{0}}$ cross sections, some resonance-like structures are seen at 
$E{_{\gamma}}\sim$~1.8 GeV and 2.8 GeV ($\sqrt{s}\sim$~2.06 GeV and 2.47 GeV) at $0.7<\cos \theta_{c.m.}^{K^+}<1.0$. 
These structures have also been observed by the CLAS Collaboration
at $\cos \theta_{c.m.}^{K^+}= 0.9$~\cite{CLAS_bradford,CLAS_dey}. 

In comparing the real data with theoretical predictions, both RPR(-Regge) and BG models fail to give a good description of the data of $K^{+}{\Lambda}$ and $K^{+}{\Sigma^{0}}$ except for the most backward bin of $\cos \theta_{c.m.}^{K^+}=
0.65$. The current new results shall enable an improvement of theoretical modeling at the forward direction for both $K^{+}{\Lambda}$ and 
  $K^{+}{\Sigma^{0}}$ channels. The structures observed at $E_{\gamma}\sim 1.8$ and 2.8 GeV ($\sqrt{s}\sim$ 2.06 and 2.47 GeV) over $0.7<\cos \theta_{c.m.}^{K^+}<1.0$ in $K^{+}{\Sigma^{0}}$ channel cannot be well reproduced by both theoretical predications.
The difference between the predictions of RPR (red solid lines) and 
RPR-Regge (blue dashed lines) indicates the contributions of nucleon 
resonances in the $s$ channel in $K^{+}{\Sigma^{0}}$ production at \eg~$<2.2$ GeV.

\begin{figure*}[htbp]
\includegraphics[width=0.8\textwidth]{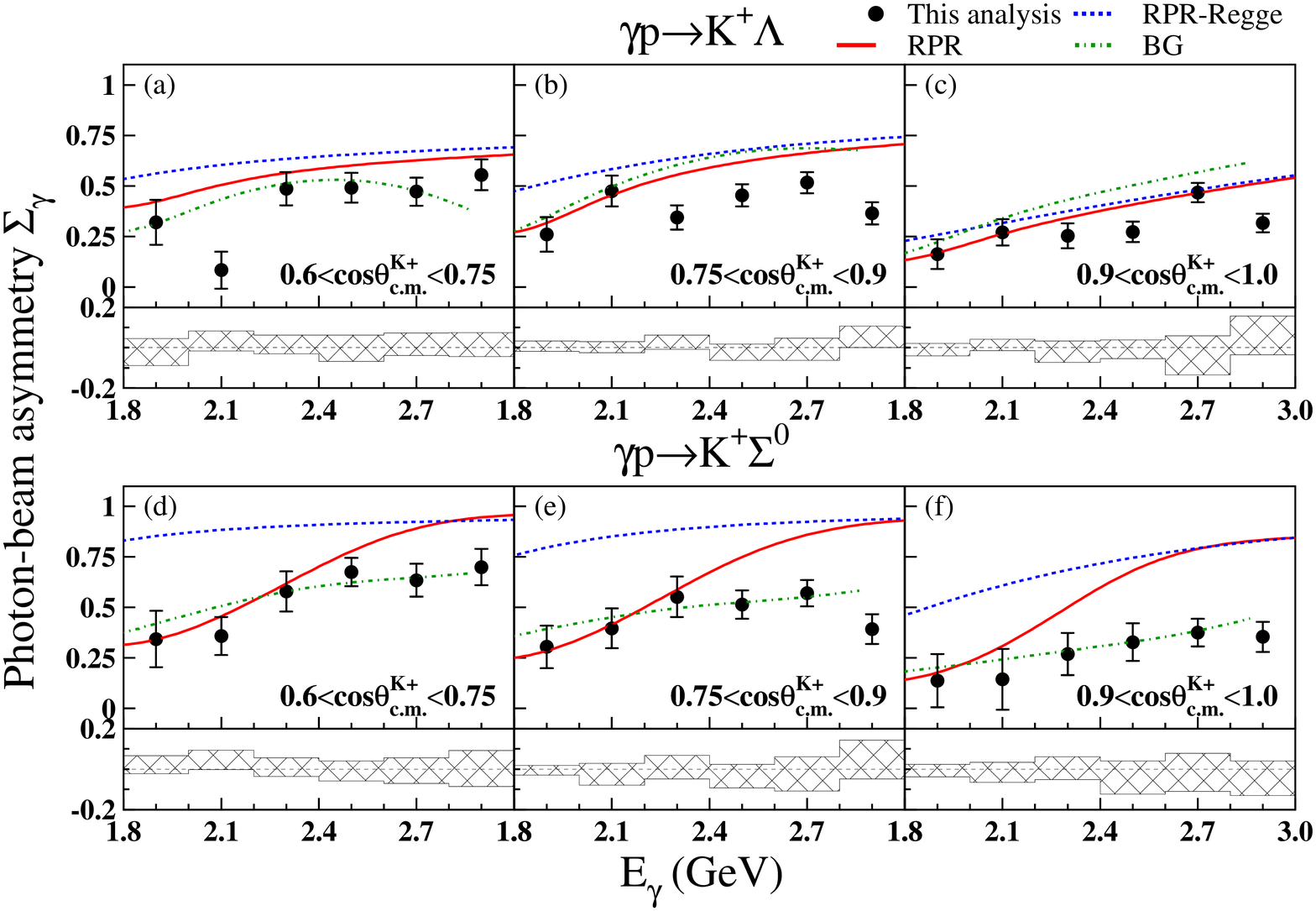}
\caption{\label{fig:beamasysys_baeg} Photon-beam asymmetries
  ($\Sigma_{\gamma}$) for the $p$($\gamma$,$K^{+}$)$\Lambda$ reaction (a)--(c) and for
  $p$($\gamma$,$K^{+}$)$\Sigma^{0}$ reaction (d)--(f) with systematic uncertainty plotted in hatched histogram as a function of photon
  energy $E_{\gamma}$ for the kaon c.m. production polar angle
  $0.6<\cos \theta_{c.m.}^{K^+}<1.0$.
The curve notations are the same as those in Fig.~\ref{fig:creg_merge_4}.
}
\end{figure*}
  
In Figs.~\ref{fig:crcos_L_4} and~\ref{fig:crcos_S_4}, the same data 
are drawn as a function of the reduced four-momentum transfer 
$|t-t_{min}|$, for 15 photon energy bins within $E_{\gamma}= 1.5$--3.0 GeV, 
where $t_{min}$ denotes $t$ at zero degrees.
Other than the theoretical predictions of RPR and BG, the 
previous results of LEPS 2006~\cite{LEPS_sumihama} (red open squares), 
SAPHIR 2004~\cite{SAPHIR_glander} (yellow open cross), CLAS 
2006~\cite{CLAS_bradford} (blue open triangles), and CLAS 
2010~\cite{CLAS_mccracken,CLAS_dey} (purple open triangles) 
are displayed as well. 
Considering the overall errors, the current 
results agree with previous measurements in overlapping kinematic 
regions. For the two sets of CLAS data~\cite{CLAS_bradford,CLAS_mccracken,CLAS_dey}, the CLAS 2010 set agrees better with our results in the $K^{+}{\Lambda}$ channel, while there is better agreement with the CLAS 2006 set in the $K^{+}{\Sigma^{0}}$ channel.

There are qualitative differences in the $t$ dependence of the differential cross sections for $K^{+}\Lambda$ and $K^{+}\Sigma^0$ at forward angles.
At low energies, the production of $K^{+}\Lambda$ shows a clear increase toward $t=t_{min}$. 
Above $E_{\gamma}>2.2$ GeV, the presence of a plateau with 
a close-to-zero slope near $t=t_{min}$ is observed for 
the $K^{+}\Lambda$ channel.
This observation is consistent with the measurements at $E_{\gamma}=5$ 
GeV~\cite{Boyarski:1969iy}. 
There even appeared a decrease of the cross 
sections toward $t=t_{min}$ for $E_{\gamma}=8$, 11 and 16 
GeV~\cite{Boyarski:1969iy}. 
As for the $K^{+}\Sigma^0$ production, the overall 
$t$ dependence is roughly flat for $E_{\gamma}<1.7$ GeV. A plateau
structure near $t=t_{min}$ with a finite negative  
slope beyond $|t-t_{min}|<0.3$ GeV$^2$ appears in $E_{\gamma}=1.7$--2.7
GeV. Going beyond $E_{\gamma}>2.7$ GeV, the $t$ dependence of the differential cross sections nears a monotonic increase toward 
$t=t_{min}$.

In Ref.~\cite{Guidal:1997hy}, the main contributions to $K^{+}\Lambda$ 
production for $E_{\gamma}>5$ GeV are described by the Reggeized 
$K^{\ast}$ exchange in the $t$ channel, and the differential cross sections have an exponential 
$t$ dependence but decrease quickly to zero at 
$t=t_{min}$. 
The plateau near $t=t_{min}$ in the differential 
cross sections at $E_{\gamma}=5$ GeV was interpreted as due to the 
contributions of a Reggeized $s$-channel diagram, which is necessitated by gauge invariance and required only for 
$K$ exchange.
Beyond the very 
forward region at $|t-t_{min}| \approx m_{K}^2$, i.e., 0.25 GeV$^2$, 
$K^{\ast}$ exchange gives the main contribution to the cross sections. 
The lack of a similar plateau feature in $K^{+}\Sigma^0$ production 
is because of a relatively minor contribution of $K$ exchange arising from
a weak coupling among $K$, $\Sigma^0$, and nucleons~\cite{Guidal:1997hy}. 

The model of contributions of a Reggeized $s$-channel
diagram for $K$ exchange at the region of $|t-t_{min}| \approx
m_{K}^2$ provides a reasonable qualitative description of what we
observe in hyperon production at $E_{\gamma}=1.5$--3.0 GeV. In
$K^{+}\Lambda$ production, we see the same plateau in
the $t$ dependence of the differential cross sections for $E_{\gamma}>2.2$
GeV. As the energy decreases, the contributions of $K$ exchange,
characterized by an increase of cross section toward $t=t_{min}$,
become more important~\cite{LEPS_sumihama}. In the
  framework of a Regge model, the energy dependence of the
  differential cross sections at $t=t_{min}$ scales as
  $s^{2\alpha_{0}-2}$ where $\alpha_0$ is the intercept of the Regge
  trajectory at $t=0$~\cite{Guidal:1997hy}. The smallness of
  $\alpha_0$ of the $K$ Regge trajectory, compared with that of $K^{\ast}$,
  would lead to an increasingly strong contribution from the $t$-channel
  $K$ exchange toward low energies. This expectation indeed agrees with what
  is observed.

For $K^{+}\Sigma^0$ production, the contributions of 
$K$ exchange are negligible overall and thus we do not observe a 
similar rise toward $t=t_{min}$ at low energies. 
Instead, the relatively flat $t$ dependence 
reflects the importance of $s$-channel nucleon resonance contributions 
in this channel. 
This can be understood because only the intermediate nucleon resonances 
with isospin $I = 1/2$ are allowed for $K^{+}\Lambda$ production, while 
both $I = 1/2$ and $3/2$ resonances could work for the case of $K^{+}\Sigma^0$. 

\begin{figure}[t]
\includegraphics[width=0.5\textwidth]{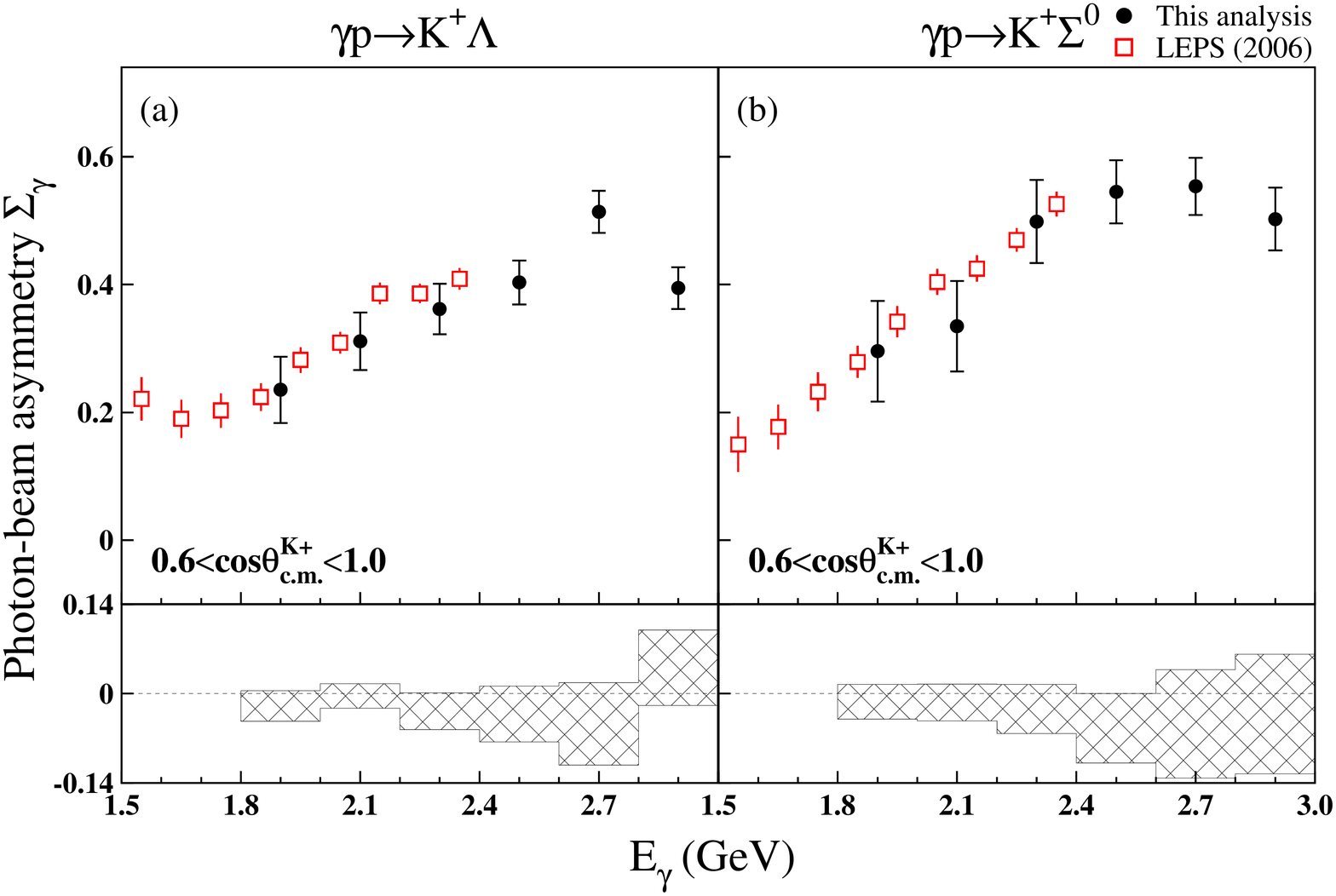}
\caption{\label{fig:beamasysys_basumcos} Photon-beam asymmetries
  ($\Sigma_{\gamma}$) for the $p$($\gamma$,$K^{+}$)$\Lambda$ reaction (a) and
  $p$($\gamma$,$K^{+}$)$\Sigma^{0}$ reaction (b) with systematic uncertainty
  plotted in hatched histogram as a function of \eg~for $0.6<\cos\theta^{K^{+}}_{c.m.}<1.0$.
  The results of this measurement and LEPS 2006 are shown.
}
\end{figure}

As for a comparison with theoretical models in the $K^{+}{\Lambda}$
channel, the RPR model describes the forward peaking rather well, except
that there exists some quantitative deviation from the data for
$E_{\gamma}>2.1$ GeV.
In the region of $|t-t_{min}|>$ 0.2 GeV$^2$, the BG model describes the data slightly
better than RPR. But the predictions of the BG model fail to reproduce the
increase at the most forward angle bin for higher energies.

In the $K^{+}{\Sigma^{0}}$ channel, the contribution of resonances is
important for $E_{\gamma}<2.2$ GeV as seen from the difference
between the predictions of RPR and RPR-Regge models. 
Both RPR and BG models fail to describe the appearance of 
forward peaking in the region of $|t-t_{min}|<$ 0.3 GeV$^2$ 
for $E_{\gamma}>2.7$ GeV. 
This brings up the need for improving the description of 
Regge trajectories for $t$-channel contributions in this energy regime. 

\subsection{Photon-beam asymmetry}
\label{sec3.2}

Figure~\ref{fig:beamasysys_baeg} shows the photon-beam asymmetries for the 
$K^{+}\Lambda$ and $K^{+}{\Sigma^{0}}$ channels as a function of 
$E_{\gamma}$ for $E_{\gamma}>1.8$ GeV in three bins of production 
angle $0.6<\cos \theta_{c.m.}^{K^+}<1.0$.
Results at very forward angles for photon energies of
2.4--3.0 GeV are obtained for the first time.
The photon-beam asymmetries are all positive and 
show a mild increase with beam energy, from $\sim$0.1--0.2 
at $E_{\gamma}=1.9$ GeV to $\sim$0.5--0.6 at $E_{\gamma}=2.9$ GeV.
In both channels, a saturation of the photon-beam asymmetries at $E_{\gamma}=2.9$ GeV in the production angle $0.75<\cos \theta_{c.m.}^{K^+}<1.0$ is observed.
In Fig.~\ref{fig:beamasysys_basumcos}, we plot the photon-beam asymmetries in the whole region of $0.6<\cos \theta_{c.m.}^{K^+}<1.0$ together with previous results at slightly lower energies~\cite{sumihama:thesis}. An increase of the photon-beam asymmetry with beam energy is more clearly illustrated.

Assuming $t$-channel dominance, based on the observed
forward-peaking feature in the differential cross sections, positive 
values of the photon-beam asymmetry suggest the dominance of natural-parity exchange 
of $K^{*}$ compared with unnatural-parity exchange of $K$ 
in the $t$ channel toward large \eg. 
Large photon-beam asymmetries are also observed 
at $E_{\gamma}=16$ GeV by SLAC~\cite{Quinn:1979zp}. 
Furthermore the photon-beam asymmetries for $K^{+}\Sigma^0$
production are slightly larger than those for $K^{+}\Lambda$
production at $E_{\gamma}>2.4$ GeV.
This indicates a relatively weaker strength of 
unnatural-parity $K$ exchange in $K^{+}\Sigma^0$ production. 
These interpretations are consistent with what we observe 
in the $t$ dependence of the differential cross sections in Sec.~\ref{sec3.1}. 

For the photon-beam asymmetry of $K^{+}\Lambda$ production below
$E_{\gamma}=2.1$ GeV and $K^{+}\Sigma^{0}$ production below
$E_{\gamma}=2.4$ GeV in Fig.~\ref{fig:beamasysys_baeg}, the inclusion of contributions from nucleon
resonances in the $s$ channel is crucial, judging from differences between data and
the predictions from Regge trajectories only (RPR-Regge). 
This feature is also found in a comparison of the production cross sections. 
For $K^{+}\Lambda$ production above $E_{\gamma}=2.1$ GeV, all predictions
from RPR, RPR-Regge, and BG converge at $\cos \theta_{c.m.}^{K^+}>0.75$ and
show certain deviations from the data.
This suggests the need for including additional resonance contributions
or a redetermination of Regge contributions with the current new data
at forward angles. 

For $K^{+}\Sigma^{0}$ production, the RPR model with no significant nucleon contributions overestimates the photon-beam 
asymmetries at $E_{\gamma}>2.3$ GeV for all three angular bins, while 
the BG model gives a very good description of the photon-beam asymmetries over
the whole region. 
It is noted that nucleon resonances with spin $J$ larger than $3/2$ 
are not included in the RPR model~\cite{RPR_2006} but they 
are taken into account in the BG model~\cite{BG_2010,BG_2011,BG_2014}. 
This difference of including higher-spin resonances might account 
for the better prediction of photon-beam asymmetries in the BG model. 

\begin{figure*}[htbp]
\includegraphics[width=0.8\textwidth]{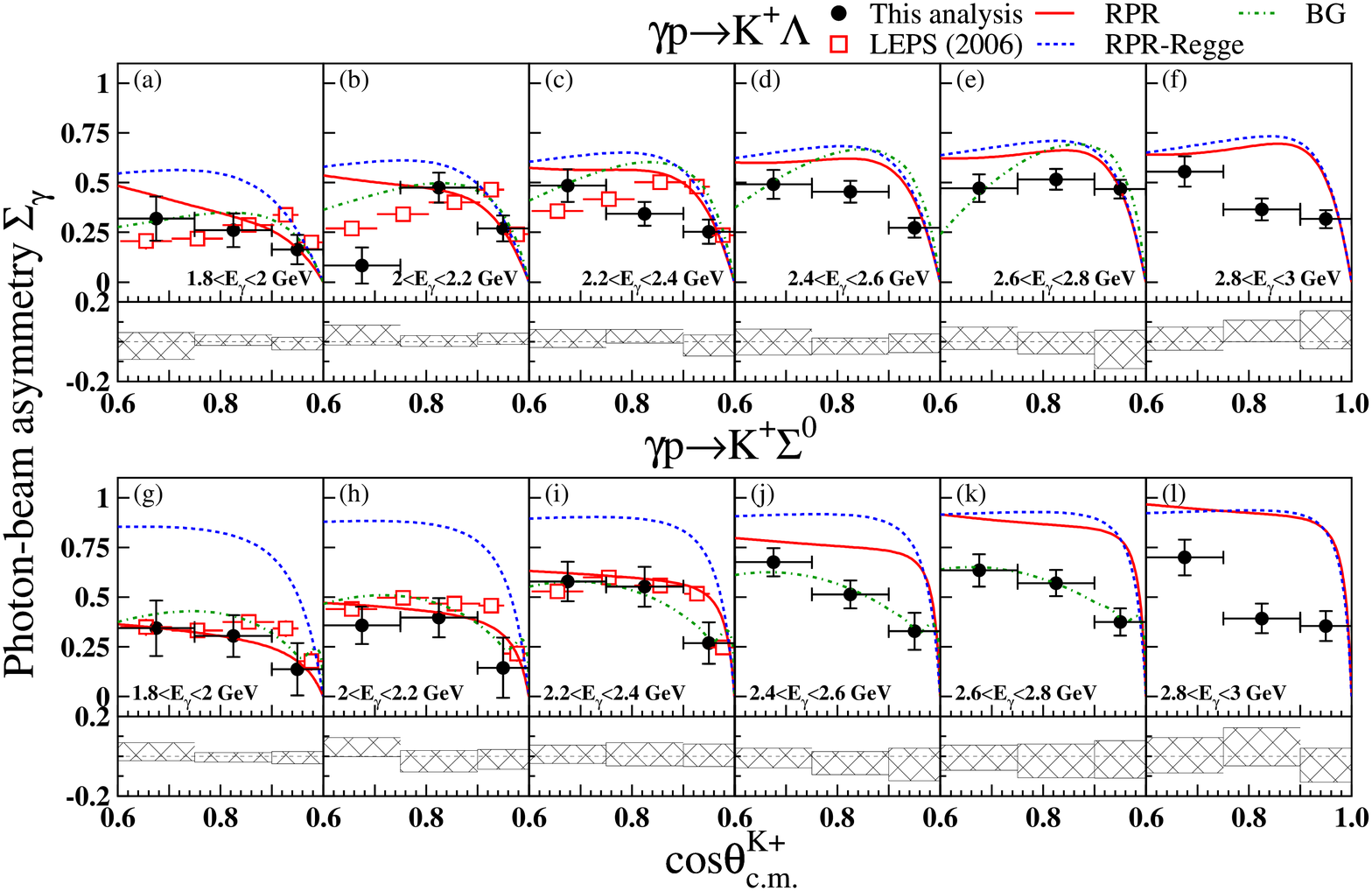}
\caption{\label{fig:beamasysys_bacos} Photon-beam asymmetries ($\Sigma_{\gamma}$) for
  the $p$($\gamma$,$K^{+}$)$\Lambda$ reaction (a)--(f)
  and $p$($\gamma$,$K^{+}$)$\Sigma^{0}$ reaction (g)--(l)
  with systematic uncertainty plotted in hatched histogram as
  a function of \cmangle for $1.8<E_{\gamma}<3.0$ GeV. The
  results of this measurement and, LEPS 2006 are shown. The curve notations are the same as those in Fig.~\ref{fig:creg_merge_4}.
}
\end{figure*}

In Fig.~\ref{fig:beamasysys_bacos}, the photon-beam asymmetry results for
$K^{+}\Lambda$ and $K^{+}{\Sigma^{0}}$ as a function of 
$\cos\theta^{K^{+}}_{c.m.}$ are shown in six \eg~bins together with the previous results from LEPS~\cite{LEPS_sumihama} 
and the theoretical predictions. The agreement with previous measurements is 
reasonably good.
Across all energy bins of \eg~from 1.8 to 3.0 GeV, the photon-beam asymmetries for the $K^{+}{\Sigma^{0}}$ channel at the forward angles of $0.6<\cos \theta_{c.m.}^{K^+}<1.0$ 
show a decrease toward zero. Such a decrease could reflect an increasing contribution of unnatural-parity $K$ exchange at smaller production angles, besides the trivial kinematic effect of a vanishing photon-beam asymmetry at zero degrees. For $E_{\gamma}<2.4$ GeV, both RPR and BG models
describe the data well and the BG model clearly does a better job in
describing the photon-beam asymmetries of $K^{+}{\Sigma^{0}}$ channel for
\eg$>2.4$ GeV.

\section{Summary}
\label{sec4}

In summary, we performed a measurement of differential cross
sections and photon-beam asymmetries for the reactions $\gamma p \rightarrow
K^{+}\Lambda$ and $K^{+}\Sigma^{0}$. 
The measured photon energy range of the
results is 1.5--3.0 GeV, with a very forward angular coverage of
$0.6<\cos \theta_{c.m.}^{K^+}<1.0$. 
The data are in good agreement with results from previous measurements~\cite{LEPS_sumihama}.

The production cross sections decrease slowly with increasing photon beam 
energy \eg. There is no observation of distinct 
narrow structures in the energy dependence of either reaction. 
For $K^{+}\Lambda$ production, there is a clear forward-peaking 
feature in the angular distributions toward large \eg, 
consistent with $t$-channel dominance for diffractive production 
at higher energies. 
The $K^{+}\Sigma^{0}$ channel shows less energy and 
angular dependence, compared to $K^{+}\Lambda$. 
This might reflect 
that $s$-channel contributions from nucleon resonances are more 
effective in the production of $K^{+}\Sigma^{0}$. 
Based on the 
Reggeized $t$-channel framework, the results of cross sections near 
$t=t_{min}$ provide evidence of the existence of $K$ exchange in
$K^{+}\Lambda$ production at low energies. 

The photon-beam asymmetries for both $K^{+}\Lambda$ and $K^{+}\Sigma^{0}$ 
channels are all positive. 
This suggests a dominating natural-parity exchange of $K^{*}$ 
in the $t$ channel. 
At \eg$>2.4$ GeV, the results deviate from the predictions of the $t$-channel 
Regge trajectories only. 
The BG model including higher-spin 
nucleon resonances describes nicely the photon-beam asymmetries for the 
$K^{+}\Sigma^{0}$ channel up to $E_{\gamma}=2.8$ GeV. All these observations strongly
suggest the existence of nucleon resonance contributions at $E_{\gamma}=2.4$--3.0 GeV.

A comparison with theoretical predictions from both RPR and BG models 
indicates that there is room for improvement of the theoretical modeling of 
Regge trajectories in the $t$ channel as well as the contributions from 
the nucleon resonances. 
With the constraints of these new data of 
hyperon $\Lambda$ and $\Sigma^{0}$ photoproduction at very forward
angles for few-GeV photons, we look forward to the
progress in theoretical modeling that will be made shortly.

\begin{acknowledgments}
The authors gratefully acknowledge the contributions of the SPring-8 
staff for supporting BL33LEP beamline and the LEPS experiment.
The experiments were performed at the BL33LEP of SPring-8 with the approval
of the Japan Synchrotron Radiation Research Institute (JASRI) as a contract
beamline (Proposal No. BL33LEP/6001). We thank T. Mart, J. Ryckebusch,
S.-H. Kim, and A. Hosaka for fruitful discussions.
This work was supported in part by the Ministry of Science and Technology of Taiwan, the Ministry of Education, 
Science, Sports and Culture of Japan, the National Research Foundation 
of Korea. and the U.S. National Science Foundation, 
\end{acknowledgments}

%\bibliography{klambda.bib}% Produces the bibliography via BibTeX.

\begin{thebibliography}{99}

\bibitem{Mart:2009nj}
  T.~Mart,
  %``Progress and Issues in the Electromagnetic Production of Kaon on the Nucleon,''
  Int.\ J.\ Mod.\ Phys.\ E {\bf 19}, 2343 (2010).

\bibitem{Klempt:2017lwq} 
  E.~Klempt, A.~V.~Sarantsev and U.~Thoma,
  %``Partial wave analysis,''
  EPJ Web Conf.\  {\bf 134}, 02002 (2017).

\bibitem{Edwards:2011jj}
  R.~G.~Edwards, J.~J.~Dudek, D.~G.~Richards, and S.~J.~Wallace,
  %``Excited state baryon spectroscopy from lattice QCD,''
  Phys.\ Rev.\ D {\bf 84}, 074508 (2011).

\bibitem{Crede:2013sze}
  V.~Crede and W.~Roberts,
  %``Progress towards understanding baryon resonances,''
  Rep.\ Prog.\ Phys.\  {\bf 76}, 076301 (2013).

\bibitem{Boyarski:1969iy}
  A.~Boyarski {\it et al.},
  %``PHOTOPRODUCTION OF K+ LAMBDA AND K+ SIGMA0 FROM HYDROGEN FROM 5-GeV to 16-Gev,''
  Phys.\ Rev.\ Lett.\  {\bf 22}, 1131 (1969).

\bibitem{Quinn:1979zp}
  D.~J.~Quinn, J.~P.~Rutherfoord, M.~A.~Shupe, D.~J.~Sherden, R.~H.~Siemann, and C.~K.~Sinclair,
  %``A Study of Charged Pseudoscalar Meson Photoproduction From Hydrogen and Deuterium With 16-{GeV} Linearly Polarized Photons,''
  Phys.\ Rev.\ D {\bf 20}, 1553 (1979).

\bibitem{SAPHIR_tran}{M.~Q.~Tran {\it et al.} (SAPHIR Collaboration), 
{Phys. Lett. B} {\bf 445}, 20 (1998).}

\bibitem{SAPHIR_glander}{K.-H.~Glander {\it et al.} (SAPHIR Collaboration), 
{Eur. Phys. J. A} {\bf 19}, 251 (2004).}

\bibitem{LEPS_zegers}{R.~G.~T.~Zegers {\it et al.} (LEPS Collaboration), 
{Phys. Rev. Lett.} {\bf 91}, 092001 (2003).}

\bibitem{LEPS_sumihama}{M.~Sumihama {\it et al.} (LEPS Collaboration), 
{Phys. Rev. C} {\bf 73}, 035214 (2006).}

\bibitem{LEPS_hicks}{K.~Hicks {\it et al.} (LEPS Collaboration), 
{Phys. Rev. C} {\bf 76}, 042201(R) (2007).}
\bibitem{LEPS_kohri}{H.~Kohri {\it et al.} (LEPS Collaboration), 
{Phys. Rev. Lett.} {\bf 97}, 082003 (2006).}

\bibitem{CLAS_bradford}{R.~Bradford {\it et al.} (CLAS Collaboration), 
{Phys. Rev. C} {\bf 73}, 035202 (2006).}

\bibitem{CLAS_mccracken}{M.~E.~McCracken {\it et al.} (CLAS Collaboration), 
{Phys. Rev. C} {\bf 81}, 025201 (2010).}

\bibitem{CLAS_dey}{B.~Dey {\it et al.} (CLAS Collaboration), 
{Phys. Rev. C} {\bf 82}, 025202 (2010).}

\bibitem{GRAAL_2007} {A.~Lleres {\it et al.} (GRAAL
  Collaboration), {Eur. Phys. J. A} {\bf 31}, 79 (2007).}

\bibitem{GRAAL_2009} {A.~Lleres {\it et al.} (GRAAL
  Collaboration), {Eur. Phys. J. A} {\bf 39}, 149 (2009).}

\bibitem{CB_Jude}{T.~C.~Jude {\it et al.} (Crystal Ball Collaboration), 
{Phys. Lett. B} {\bf 735}, 112 (2014).}

%\bibitem{LEPS_kohri}{H.~Kohri {\it et al.} (LEPS Collaboration), 
%{Phys. Rev. Lett.} {\bf 97}, 082003 (2006).}

\bibitem{CLAS_Pereira}
  S.~A.~Pereira {\it et al.} (CLAS Collaboration),
  %``Differential cross section of gamma n to K+ Sigma- on bound neutrons with incident photons from 1.1 to 3.6 GeV,''
  Phys.\ Lett.\ B {\bf 688}, 289 (2010).

\bibitem{SAPHIR_goers} S.~Goers {\it et al.} (SAPHIR Collaboration),
  %``Measurement of gamma p --> K0 Sigma+ at photon energies up to 1.55-GeV,''
  Phys.\ Lett.\ B {\bf 464}, 331 (1999).

\bibitem{SAPHIR_lawall} R.~Lawall {\it et al.} (SAPHIR Collaboration),
  %``Measurement of the reaction gamma p ---> K0 Sigma+ at photon energies up to 2.6-GeV,''
  Eur.\ Phys.\ J.\ A {\bf 24}, 275 (2005).

\bibitem{TAPS_2008}
  R.~Castelijns {\it et al.} (CBELSA/TAPS Collaboration),
  %``Nucleon resonance decay by the K0 Sigma+ channel,''
  Eur.\ Phys.\ J.\ A {\bf 35}, 39 (2008).

\bibitem{TAPS_2012}
  R.~Ewald {\it et al.} (CBELSA/TAPS Collaboration),
  %``Anomaly in the $K^0_S \Sigma^+$ photoproduction cross section off the proton at the $K^*$ threshold,''
  Phys.\ Lett.\ B {\bf 713}, 180 (2012).

\bibitem{A2}
  P.~Aguar-Bartolome {\it et al.} (A2 Collaboration),
  %``Measurement of the $\gamma p \to K^{0} \Sigma^{+}$ reaction with the Crystal Ball/TAPS detectors at the Mainz Microtron,''
  Phys.\ Rev.\ C {\bf 88}, 044601 (2013).


\bibitem{KMAID_1999}{T.~Mart and C.~Bennhold, {Phys. Rev. C} {\bf 61}, 012201(R) (1999).}

\bibitem{KMAID_2014}
  T.~Mart,
  %``Electromagnetic production of $K\Sigma$ on the nucleon near threshold,''
  Phys.\ Rev.\ C {\bf 90}, 065202 (2014).

\bibitem{KMAID_2015}
  T.~Mart, S.~Clymton and A.~J.~Arifi,
  %``Nucleon resonances with spin 3/2 and 5/2 in the isobar model for kaon photoproduction,''
  Phys.\ Rev.\ D {\bf 92}, 094019 (2015).

\bibitem{KMAID_2017}
  T.~Mart and S.~Sakinah,
  %``Multipoles model for $K^+ \Lambda$ photoproduction on the nucleon reexamined,''
  Phys.\ Rev.\ C {\bf 95}, 045205 (2017).

\bibitem{Guidal:1997hy}
  M.~Guidal, J.~M.~Laget and M.~Vanderhaeghen,
  %``Pion and kaon photoproduction at high-energies: Forward and intermediate angles,''
  Nucl.\ Phys.\ A {\bf 627}, 645 (1997).

\bibitem{Guidal:2003qs}
  M.~Guidal, J.~M.~Laget and M.~Vanderhaeghen,
  %``Exclusive electromagnetic production of strangeness on the nucleon: Review of recent data in a Regge approach,''
  Phys.\ Rev.\ C {\bf 68}, 058201 (2003).

\bibitem{BG_2005}
  A.~V.~Sarantsev, V.~A.~Nikonov, A.~V.~Anisovich, E.~Klempt and U.~Thoma,
  %``Decays of baryon resonances into Lambda K+, Sigma0 K+ and Sigma+ K0,''
  Eur.\ Phys.\ J.\ A {\bf 25}, 441 (2005).

\bibitem{BG_2007}
  A.~V.~Anisovich, V.~Kleber, E.~Klempt, V.~A.~Nikonov, A.~V.~Sarantsev and U.~Thoma,
  %``Baryon resonances and polarization transfer in hyperon photoproduction,''
  Eur.\ Phys.\ J.\ A {\bf 34}, 243 (2007).

\bibitem{BG_2010}
  A.~V.~Anisovich, E.~Klempt, V.~A.~Nikonov, A.~V.~Sarantsev and U.~Thoma,
  %``P-wave excited baryons from pion- and photo-induced hyperon production,''
  Eur.\ Phys.\ J.\ A {\bf 47}, 27 (2011).

\bibitem{BG_2011}
  A.~V.~Anisovich, E.~Klempt, V.~A.~Nikonov, A.~V.~Sarantsev and U.~Thoma,
  %``Nucleon resonances in the fourth resonance region,''
  Eur.\ Phys.\ J.\ A {\bf 47}, 153 (2011).

\bibitem{BG_2014}
  E.~Gutz {\it et al.} (CBELSA/TAPS Collaboration),
  %``High statistics study of the reaction $\gamma p\to p\pi^0\eta$,''
  Eur.\ Phys.\ J.\ A {\bf 50}, 74 (2014).

\bibitem{RPR_2006}{T.~Corthals, J.~Ryckebusch, and T. van Cauteren, 
{Phys. Rev. C} {\bf 73}, 045207 (2006).}

\bibitem{RPR_2007}{T.~Corthals, T.~van Cauteren, J.~Ryckebusch, and D. G.~Ireland, {Phys. Rev. C} {\bf 75}, 045204 (2007).}

\bibitem{RPR_2012}{ L.~De~Cruz, J.~Ryckebusch, T. Vrancx, and P. Vancraeyveld,
{Phys. Rev. C} {\bf 86}, 015212 (2012).}

\bibitem{Anisovich:2017pmi}
  A.~V.~Anisovich {\it et al.},
  %``Strong Evidence for Nucleon Resonances near 1900????MeV,''
  Phys.\ Rev.\ Lett.\  {\bf 119}, 062004 (2017).

\bibitem{muramatsu}{N.~Muramatsu {\it et al.}, 
{Nucl. Instrum. Methods A} {\bf 737}, 184 (2014).} 

\bibitem{hwang}{S.~H.~Hwang {\it et al.} (LEPS Collaboration), 
{Phys. Rev. Lett.} {\bf 108}, 092001 (2012).}

\bibitem{nakano}{T.~Nakano {\it et al.} (LEPS Collaboration), 
{Nucl. Phys. A} {\bf 684}, 71 (2001).}

\bibitem{RPR_url}\url{http://rprmodel.ugent.be/calc/}

\bibitem{BG_url}\url{http://pwa.hiskp.uni-bonn.de/BG2014_02_obs_int.htm}

\bibitem{sumihama:thesis}{M.~Sumihama, Ph.D. thesis, Osaka University, 2003 (unpublished).}

%%%%%%%%%%%%%%%%%%%%%%%%%%%%%%%%%%%%%%%%%%%%%%%%%%%%%%%%%%%%%%%%%%%%%%

\end{thebibliography}

\end{document}